\documentclass[10pt,conference]{IEEEtran}
\IEEEoverridecommandlockouts
\usepackage{cite}
\usepackage{caption}
\usepackage{amsmath,amssymb,amsfonts}
\usepackage{algorithmic}
\usepackage{graphicx}
\usepackage{booktabs}
\usepackage{bm}
\usepackage{setspace}
\usepackage{textcomp}
\usepackage{xcolor}
\usepackage{dblfloatfix}
\usepackage{amsmath}
\usepackage{subfig}
\usepackage{float} 
\usepackage{array}
\usepackage{calc}  
\usepackage{enumitem}
\fontsize{10pt}{12pt}\selectfont
\usepackage{multirow}
\graphicspath{{pictures/}}
\begin{document}

\title{Cascaded channel modeling and experimental validation for RIS assisted communication system\\
}

\author{\IEEEauthorblockN{Jiwei Zhang\IEEEauthorrefmark{1},  Yuxiang Zhang\IEEEauthorrefmark{1}, Tao Jiang\IEEEauthorrefmark{2}, Huiwen Gong\IEEEauthorrefmark{1}, Hongbo Xing\IEEEauthorrefmark{1}, and Lei Tian\IEEEauthorrefmark{1}}

\IEEEauthorblockA{\IEEEauthorrefmark{1}\textit{State Key Lab of Networking and Switching Technology, Beijing University of Posts and Telecommunications, Beijing, China} \\}
\IEEEauthorblockA{\IEEEauthorrefmark{2}\textit{China Mobile Research Institution, Beijing, China} \\}
\IEEEauthorblockA{\IEEEauthorrefmark{1}\{rediaose, zhangyx, birdsplan, hbxing, tianlbupt\}@bupt.edu.cn \\}
\IEEEauthorblockA{\IEEEauthorrefmark{2}jiangtao@chinamobile.com}
}

\maketitle

\begin{abstract}
Reconfigurable Intelligent Surface (RIS) is considered as a promising technology for 6G due to its ability to actively modify the electromagnetic propagation environment. Accurate channel modeling is essential for the design and evaluation of RIS assisted communication systems. Most current research models the RIS channel as a cascade of Tx-RIS and RIS-Rx sub-channels. However, most validation efforts regarding this assumption focus on large-scale path loss. To further explore this, in this paper, we derive and extend a convolution expression of RIS cascaded channel model based on the previously proposed Geometry-based Stochastic Model (GBSM)-based RIS cascaded channels. This model follows the 3GPP standard framework and leverages parameters such as angles, delays, and path powers defined in the GBSM model to more accurately reflect the small-scale characteristics of RIS multipath cascades. To verify the accuracy of this model, we conduct measurements of the Tx-RIS-Rx channel, Tx-RIS, and RIS-Rx sub-channels in a factory environment at 6.9 GHz, using the measured data to demonstrate the model's validity and applicability in real-world scenarios. Validation with measured data shows that the proposed model accurately describes the characteristics of the RIS cascaded channel in terms of delay, angle, and power in complex multipath environments, providing important references for the design and deployment of RIS systems.
\end{abstract}


\section{Introduction}
Reconfigurable Intelligent Surfaces (RIS) consists of a two-dimensional metasurface composed of numerous sub-wavelength passive reflecting elements. By adjusting the reflection or transmission coefficients of each unit, RIS actively modifies the electromagnetic propagation environment, enabling use cases such as enhanced signal coverage, smart city applications, and Internet of Things (IoT) communications \cite{ctj}, therefore being considered a promising technology for 6G applications \cite{basar}.

Accurate channel modeling is crucial for the design and evaluation of RIS communication systems \cite{zongshu}. In the study of RIS channel models, the main difference from traditional channel models lies in the introduction of the Tx-RIS-Rx channel. This paper refers to this as the RIS channel, which is typically modeled as a cascade of the Tx-RIS and RIS-Rx sub-channels. Many studies on RIS communication system design are based on this cascaded model assumption \cite{mtx1}, \cite{mtx2}. Several works have validated this cascaded model assumption through simulations and measurements.

In terms of simulation validation, some studies validate this assumption through Ray Tracing (RT) simulations from multiple parameters. \cite{wcx} proposes a segmented simulation method based on the cascade hypothesis, performing RT  simulations on the RIS cascade channel, analyzing its path loss and angle power spectrum, and validating it against measurement results. \cite{zzf} suggests a method to model RIS response using radiation patterns and conducts cascade RT simulations, analyzing the received power variations with distance and angle. In terms of measurements validation, the empirical validation of this assumption currently mainly focuses on the cascaded characteristics of large-scale parameters such as path loss. \cite{twk} derives the free-space path loss formula for the RIS channel and verifies the model's accuracy through measurements in a microwave anechoic chamber. \cite{ly} further considers the multipath effect and measures the path loss of the RIS channel in a corridor scenario, proposing a multiplicative distance model that accounts for waveguide effects to more accurately model the path loss of the RIS channel. \cite{shibie} focuses on path loss in RIS-assisted radio frequency identification scenarios, assuming identical sub-channels for Tx-RIS and RIS-Rx, and verifies through measurements that the path loss is proportional to the fourth power of the distance. 
\begin{figure*}[t]
    \begin{align}
        h^{u, s}_{\text{Tx-RIS-Rx}}(\tau) &= \sum_{n_1,m_1}^{N_1,M_1} \sum_{n_2,m_2}^{N_2,M_2}\sqrt{P_{n_1,m_1}P_{n_2,m_2}} F_u(\theta^{\text{Rx}}_{n_2,m_2}) F_{\text{RIS}}(\theta^{\text{out}}_{n_2,m_2},  \theta^{\text{in}}_{n_1,m_1}) F_s(\theta^{\text{Tx}}_{n_1,m_1}) \nonumber \\
        & \quad \cdot \text{e}^{\text{j}\frac{2\pi}{\lambda} (|\textbf{d}_u^{\text{Rx}}| \cos \theta^{\text{Rx}}_{n_2,m_2} + |\textbf{d}_s^{\text{Tx}}| \cos \theta^{\text{Tx}}_{n_1,m_1})} \delta(\tau - \tau_{n_2,m_2} - \tau_{n_1,m_1}) \tag{1} \\
        &= \iint_{-\infty}^{+\infty}\sum_{n_1,m_1}^{N_1,M_1} \sum_{n_2,m_2}^{N_2,M_2} \int_{-\infty}^{+\infty} \sqrt{P_{n_2,m_2}} F_u(\theta^{\text{Rx}}) \delta(\theta^{\text{Rx}} - \theta_{n_2,m_2}^{\text{Rx}})\delta(\theta^{\text{out}} - \theta_{n_1,m_1}^{\text{out}}) \text{e}^{\text{j}\frac{2\pi}{\lambda} |\textbf{d}_s^{\text{Rx}}| \cos \theta^{\text{Rx}}} \text{d}\theta^{\text{Rx}} \nonumber \\
        & \quad \cdot  F_{\text{RIS}}(\theta^{\text{out}}, \theta^{\text{in}}) \cdot \int_{-\infty}^{+\infty} \sqrt{P_{n_1,m_1}} F_s(\theta^{\text{Tx}}) \delta(\theta^{\text{Tx}} - \theta_{n_1,m_1}^{\text{Tx}})\delta(\theta^{\text{in}} - \theta_{n_1,m_1}^{\text{in}}) \text{e}^{\text{j}\frac{2\pi}{\lambda} |\textbf{d}_u^{\text{Tx}}| \cos \theta^{\text{Tx}}} \text{d}\theta^{\text{Tx}} \nonumber \\
        & \quad \cdot \delta(\tau - \tau_{n_2,m_2}) * \delta(\tau - \tau_{n_1,m_1}) \text{d}\theta^{\text{in}} \text{d}\theta^{\text{out}}
        \tag{2}
    \end{align}
    \vspace{-5mm} 
    \begin{align}
        h^u_{\text{RIS-Rx}}(\theta_{\text{out}}, \tau) &= \int_{-\infty}^{+\infty} \sum_{n_2,m_2}^{N_2,M_2}  \sqrt{P_{n_2,m_2}} F_u(\theta^{\text{Rx}}) \delta(\theta^{\text{Rx}} - \theta_{n_2,m_2}^{\text{Rx}})\delta(\theta^{\text{out}} - \theta_{n_2,m_2}^{\text{out}})\delta(\tau - \tau_{n_2,m_2}) \text{e}^{\text{j}\frac{2\pi}{\lambda} |\textbf{d}_u^{\text{Rx}}| \cos \theta^{\text{Rx}}}  \text{d}\theta^{\text{Rx}}, \tag{3}
    \end{align}
    \vspace{-5mm} 
    \begin{align}
        h^s_{\text{Tx-RIS}}(\theta_{\text{in}}, \tau) &= \int_{-\infty}^{+\infty} \sum_{n_1,m_1}^{N_1,M_1}  \sqrt{P_{n_1,m_1}} F_s(\theta^{\text{Tx}}) \delta(\theta^{\text{Tx}} - \theta_{n_1,m_1}^{\text{Tx}})\delta(\theta^{\text{in}} - \theta_{n_1,m_1}^{\text{in}})\delta(\tau - \tau_{n_1,m_1}) \text{e}^{\text{j}\frac{2\pi}{\lambda} |\textbf{d}_s^{\text{Tx}}| \cos \theta^{\text{Tx}}}  \text{d}\theta^{\text{Tx}}, \tag{4}
    \end{align}
    \vspace{-5mm} 
    \begin{align}
    h^{u, s}_{\text{Tx-RIS-Rx}}(\tau) 
    & = \iint_{-\infty}^{\infty} h^u_{\text{RIS-RX}}(\theta^{\text{out}}, \tau) * h^s_{\text{TX-RIS}}(\theta^{\text{in}}, \tau) \cdot F_{\text{RIS}}(\theta^{\text{out}}, \theta^{\text{in}}) \, \text{d}\theta^{\text{out}} \, \text{d}\theta^{\text{in}}. \tag{5}
    \end{align}
\end{figure*}

Although significant progress has been made in researching RIS channel cascaded models, there is a lack of empirical validation for the cascaded relationship between multipath components of the RIS channel and sub-channels. To address this, this paper extends the Geometry-based Stochastic Model (GBSM)-based RIS channel model \cite{ghw} by presenting a convolutional form to describe the cascaded relationship between sub-channel multipaths. This model follows the 3GPP standard framework, and uses parameters like angles, delays, and path powers from the GBSM model to better represent the small-scale features of RIS multipath cascades. To verify the accuracy of this model, a channel measurement is conducted at 6.9 GHz in a factory setting, where the Tx-RIS-Rx channel and the Tx-RIS and RIS-Rx sub-channels are measured. Based on this measurement data, it is found that the proposed model fits well with the empirical results in terms of angle, delay, and power.

The remainder of the paper is organized as follows. Section II introduces a RIS channel cascaded model under the convolutional form. Section III describes the measurement system and scheme. Section IV presents the measurement results and analysis of the Power Delay Profile (PDP). Section V provides the conclusions of this study.

\section{RIS cascaded channel model under the Convolutional Form}
\begin{figure}[ht]
    \centering
    \includegraphics[width=\columnwidth]{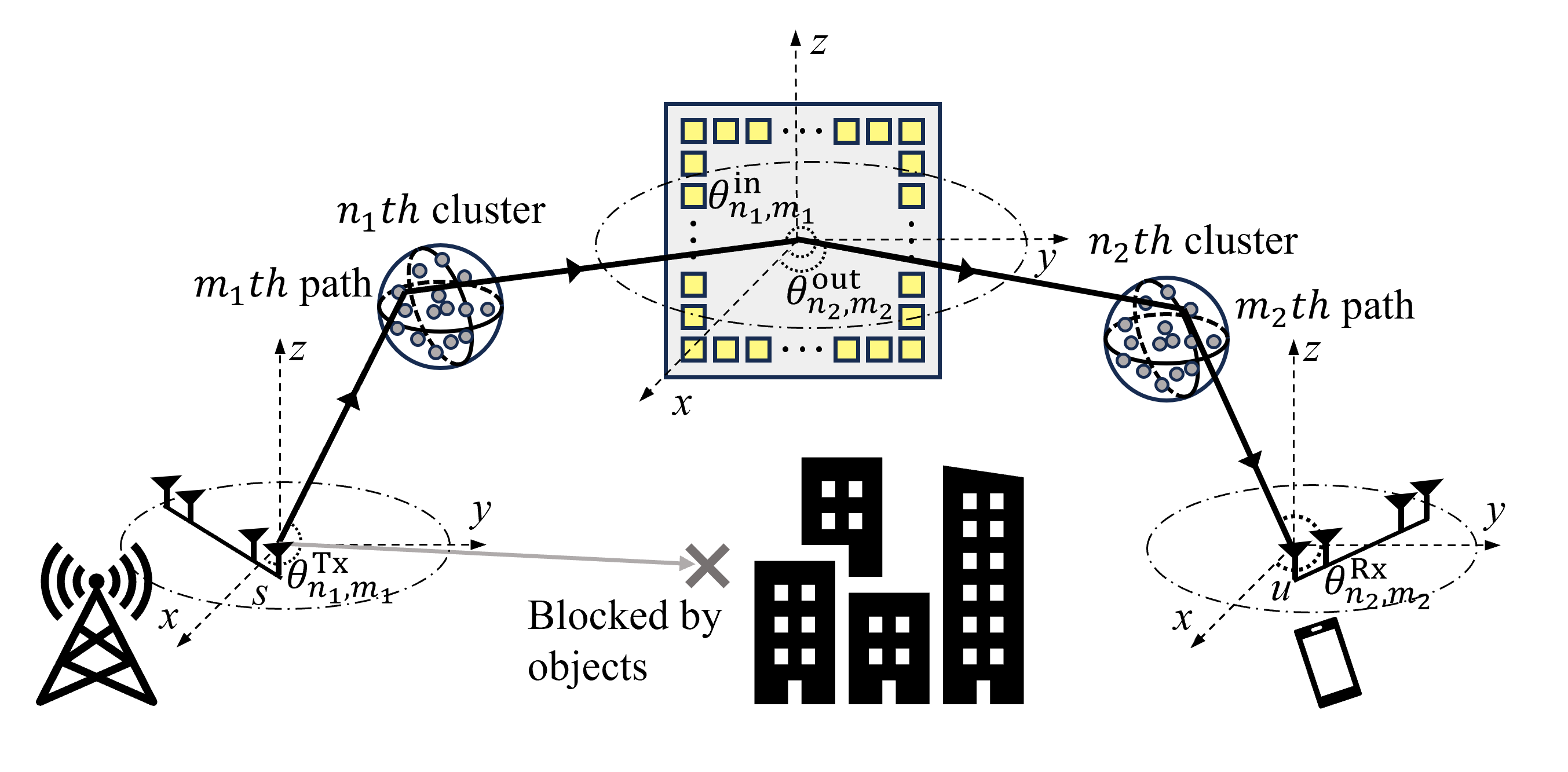}
    \caption{The system model of RIS-assisted Multiple Multiple-Input Multiple-Output (MIMO) communication. The direct link between Tx and Rx is blocked by objects \cite{ghw}.}
    \label{risModel}
\end{figure}

Consider a GBSM-based RIS channel model from \cite{ghw}, as shown in Fig.~\ref{risModel}. There are randomly distributed scatterer areas (clusters) between Tx and RIS \cite{cluster}, and between RIS and Rx, which generate a set of paths with similar attributes, assuming that the direct link between the transmitter and receiver is blocked.

The channel impulse response (CIR) of the RIS-assisted communication channel between $s$ and $u$ can be expressed as (1). To explicitly illustrate the cascading relationship between sub-channels, we reconstruct (1) into (2), where

\begin{itemize}
    \item[\textbullet] $n_1$, $n_2$, and $N_1$, $N_2$ represent the indexes and total numbers of clusters in Tx-RIS sub-channel and RIS-Rx sub-channel, respectively.
    \item[\textbullet] $m_1$, $m_2$, and $M_1$, $M_2$ denote the indexes and total numbers of the paths within the corresponding clusters.
    \item[\textbullet] $\delta(\cdot)$ denotes the Dirac Delta function.
    \item[\textbullet] $P_{n_1,m_1}$ and $P_{n_2,m_2}$ represent the power of the corresponding path, expressed as a relative value, which is the ratio of the received power of that path component to the transmitted power.
    \item[\textbullet] $\textbf{d}_u^{\text{Tx}}$ and $\textbf{d}_u^{\text{Rx}}$ indicate the location vectors of antenna $s$ and $u$.
    \item[\textbullet] $\lambda$ is the wavelength of the carrier frequency.
    \item[\textbullet] $F_{s}$, $F_{u}$, $F_{\text{RIS}}$ indicate the radiation pattern of antenna $s$ at Tx, antenna $u$ at Rx and RIS, respectively.
    \item[\textbullet] $\theta_{n_1,m_1}^{\text{in}}$ and $\theta_{n_1,m_1}^{\text{Tx}}$ denote the angle of the $(n_1, m_1)^{th}$ path in Tx-RIS sub-channel, which are the azimuth angle of arrival (AoA) and the azimuth angle of departure (AoD) respectively. Similarly, $\theta_{n_2,m_2}^{\text{Rx}}$ and $\theta_{n_2,m_2}^{\text{out}}$ are the AoA, AoD of the $(n_2, m_2)^{th}$ path in RIS-Rx sub-channel.
    \item[\textbullet] $\tau_{n_1,m_1}$ and $\tau_{n_2,m_2}$ represent the delay of the corresponding path.
\end{itemize}

 It is noteworthy that this model simplifies the consideration of vertical angles, polarization, and Doppler effects compared to 3GPP TR 38.901 \cite{38901}. This simplification allows for a focused exploration of the cascading characteristics within the RIS. However, the model remains applicable even when incorporating or extending these factors. According to (2), we define the impulse responses of the Tx-RIS and RIS-Rx sub-channels as shown in (3) and (4), respectively. The impulse response of the RIS channel can be represented as the convolution of the impulse responses of the Tx-RIS and RIS-Rx sub-channels, with the addition of the effect of the RIS radiation pattern gain, as shown in (5).

From (5), it is observed that the impact of the RIS on clusters and paths is encapsulated in the equivalent RIS radiation pattern $F_{\text{RIS}}$, which is a function of the AoA and AoD of RIS. The detailed calculation method for $F_{\text{RIS}}$ can be referred to in \cite{ghw}. 

\section{Measurement Description}
\subsection{Measurement System}
In this measurement experiment, the Tx utilizes a vector signal generator to generate a PN code sequence with a length of 511 bits. The generated signal is then modulated to 6.9 GHz using binary phase shift keying and transmitted via the transmitting antenna. At the Rx end, a spectrum analyzer is used to capture and record the received signal. A RIS is placed in the channel between TX and RX. The RIS operates at a center frequency of 6.9 GHz, with physical dimensions of 0.61 m × 0.61 m. Each row and column of the RIS consists of 32 elements arranged at half-wavelength spacing, and each element can independently control its phase with 1-bit resolution.

\begin{figure}[ht]
    \centering
    \includegraphics[width=0.9\columnwidth]{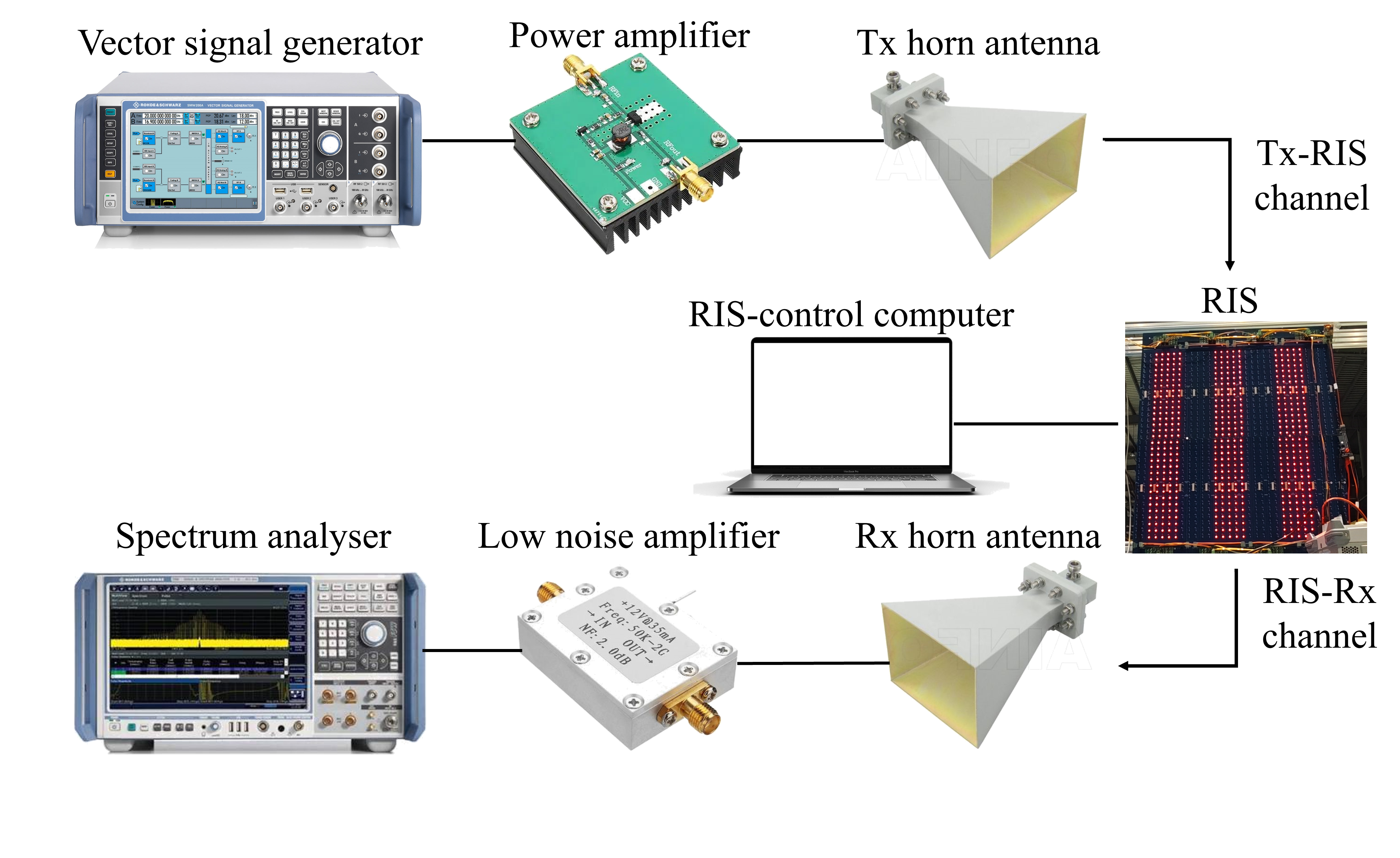}
    \caption{Measurement system diagram.}
    \label{meaSys}
\end{figure}

To ensure sufficient system gain, a low-noise amplifier is used at the receiver to amplify the signal. Fig.~\ref{meaSys} illustrates the structure of the measurement system. The specific configuration of the measurement system is detailed in Table~\ref{meaPara}.

\begin{table}[ht]
\caption{Measurement System Parameters}
\centering
\renewcommand{\arraystretch}{1.5}
\small
\tabcolsep=0.7cm
\begin{tabular}{c|c}
\hline
\hline
\textbf{Parameter} & \textbf{Value} \\ \hline
Center frequency (GHz) & 6.9 \\ \hline
Bandwidth (MHz) & 400 \\ \hline
PN sequence & 511 \\ \hline
Transmit power (dBm) & 0 \\ \hline
Horn antenna azimuth HPBW (deg) & 15 \\ \hline
Horn antenna gain (dBi) & 20 \\ \hline
Omni-directional antenna gain (dBi) & 3 \\ \hline
Antenna / RIS height (m) & 1.5 \\ \hline
Antenna polarization & H-to-H \\ \hline
\hline
\end{tabular}
\label{meaPara}
\end{table}

\subsection{Measurement Setting}
\begin{figure}[ht]
    \centering
    \includegraphics[width=0.9\columnwidth]{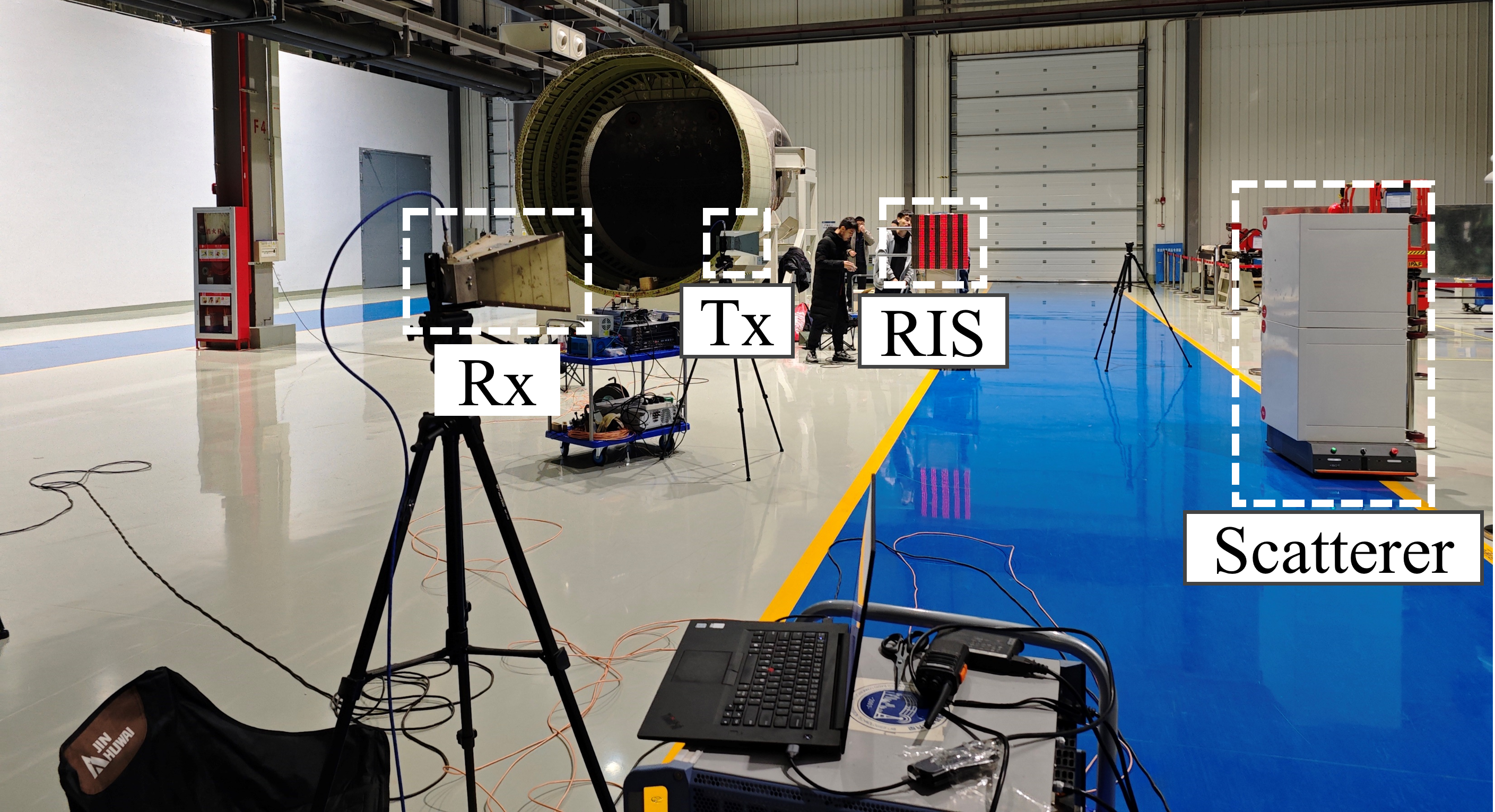}
    \caption{The measurement campaign.}
    \label{meaPhoto}
\end{figure}

As shown in Fig.~\ref{meaPhoto}, the measurements are conducted indoors in a factory enviroment. To explore the coupling relationship between the cascaded channel and sub-channels in terms of path, we measure the Tx-RIS-Rx cascaded channel, the Tx-RIS sub-channel, and the RIS-Rx sub-channel separately.

\subsubsection{Tx-RIS-Rx measurement}
\begin{figure}[ht]
    \centering
    \includegraphics[width=0.75\columnwidth]{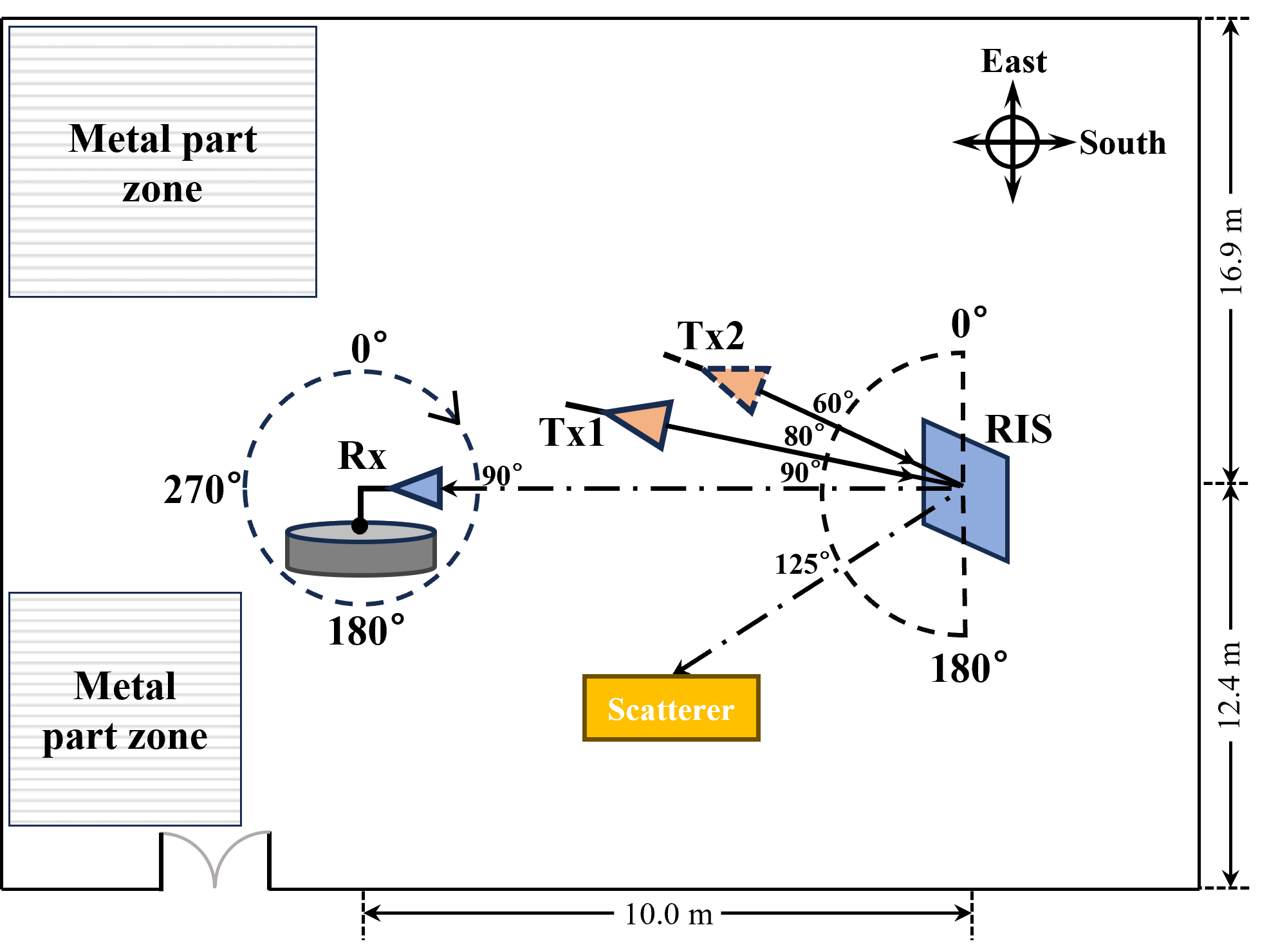}
    \caption{Schematic diagram of the Tx-RIS-Rx cascaded channel measurement Scenario.}
    \label{Tx-RIS-RxScene}
\end{figure}

The specific layout of Tx-RIS-Rx cascaded channel measurement is illustrated in Fig.~\ref{Tx-RIS-RxScene}. To facilitate subsequent analysis, the AoA or AoD at RIS are defined in the counterclockwise direction as positive, while the AoA at Rx is defined in the clockwise direction as positive. Both are referenced from the east direction as 0 degrees. The transmitting horn antenna is placed sequentially at two different positions, labeled Tx1 and Tx2, directed toward the RIS to simulate two LOS paths. At the Tx1 position, the azimuth angle relative to the RIS is 80 degrees, with a distance of 5 meters; at the Tx2 position, the azimuth angle is 60 degrees, with a distance of 3 meters. The Rx is fixed on a turntable 10 meters away from the RIS, and it rotates 360 degrees in 5-degree increments to measure the angle of arrival of the signal. To introduce multipath, a scatterer is artificially placed in the RIS-Rx sub-channel at an angle of 125 degrees to the RIS.

During the experiment, the RIS operates in an anomalous reflection mode. Based on the positions of the Tx and Rx, we calculate four sets of codebooks, referred to as codebook1, codebook2, codebook3, and codebook4. It is important to note that the experiment does not focus on the specific directions in which these codebooks perform anomalous reflections. The primary purpose of using multiple codebooks is to verify the validity of the experiment and the stability of the results by altering the RIS radiation pattern. Additionally, for comparative analysis, measurements are also taken with the RIS removed.

\subsubsection{Tx-RIS measurement}
During the Tx-RIS sub-link measurement, the transmitting horn antenna is placed in the same position as in the cascaded link measurement, while an omni-directional antenna is placed at the RIS to measure the signal power and delay of the Tx-RIS segment.

\subsubsection{RIS-Rx measurement}
\begin{figure}[!htbp]
    \centering
    \subfloat[]{\includegraphics[width=0.5\columnwidth]{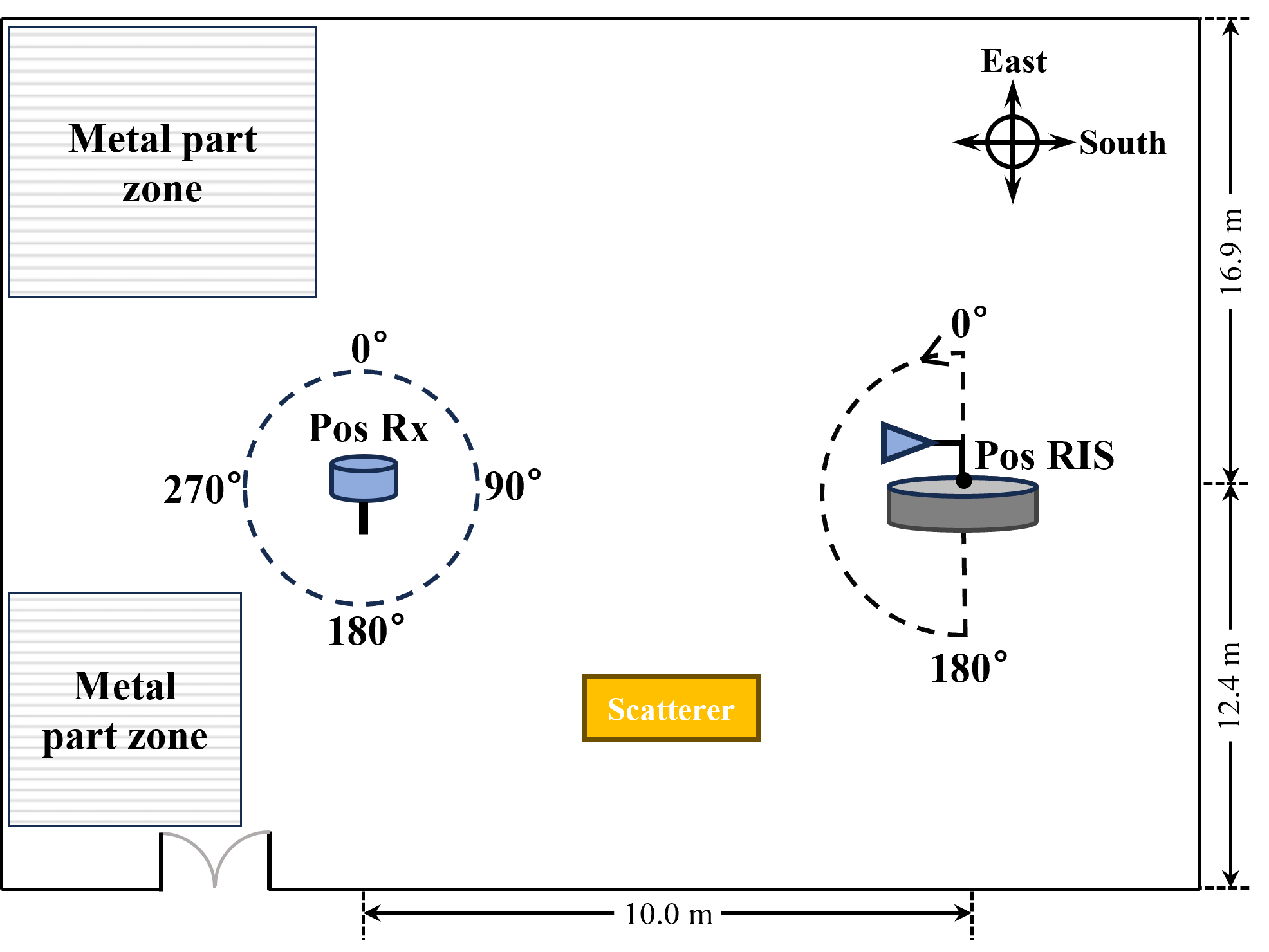}\label{RisDep}}
    \hfill
    \subfloat[]{\includegraphics[width=0.5\columnwidth]{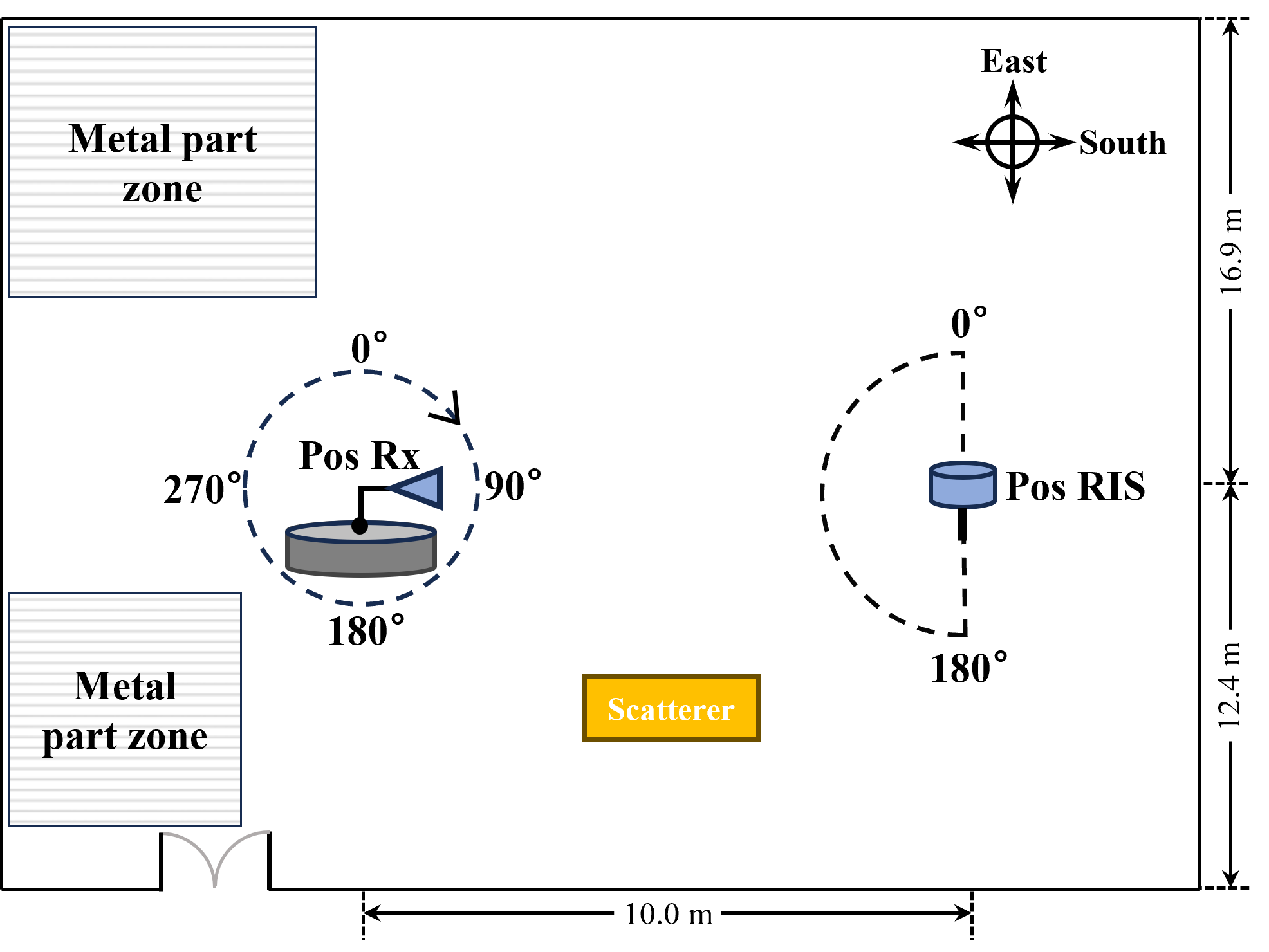}\label{RxArr}}
    \caption{Schematic diagram of RIS-Rx sub-channel measurement, (a) measurement of the AoD of paths from the RIS, (b) measurement of the AoA of paths at the Rx.}
    \label{RIS-Rx}
\end{figure}

For the RIS-Rx sub-channel, accurately obtaining the departure angles of paths from the RIS is essential to calculate the RIS gain. To achieve this, we design two sets of experiments, as shown in Fig.~\ref{RIS-Rx}, to measure the RIS departure angles and the Rx arrival angles of the paths. In Fig.~\ref{RIS-Rx}(a), a horn antenna is fixed on a rotating platform at the RIS location and rotated 180 degrees in 5-degree increments to measure the RIS departure angles, with an omni-directional antenna placed at the Rx side. In Fig.~\ref{RIS-Rx}(b), a horn antenna at the Rx is rotated 360 degrees to measure the Rx arrival angles, with an omni-directional antenna placed at the RIS side.

\begin{figure}[!htbp]
    \centering
    \includegraphics[width=0.9\columnwidth]{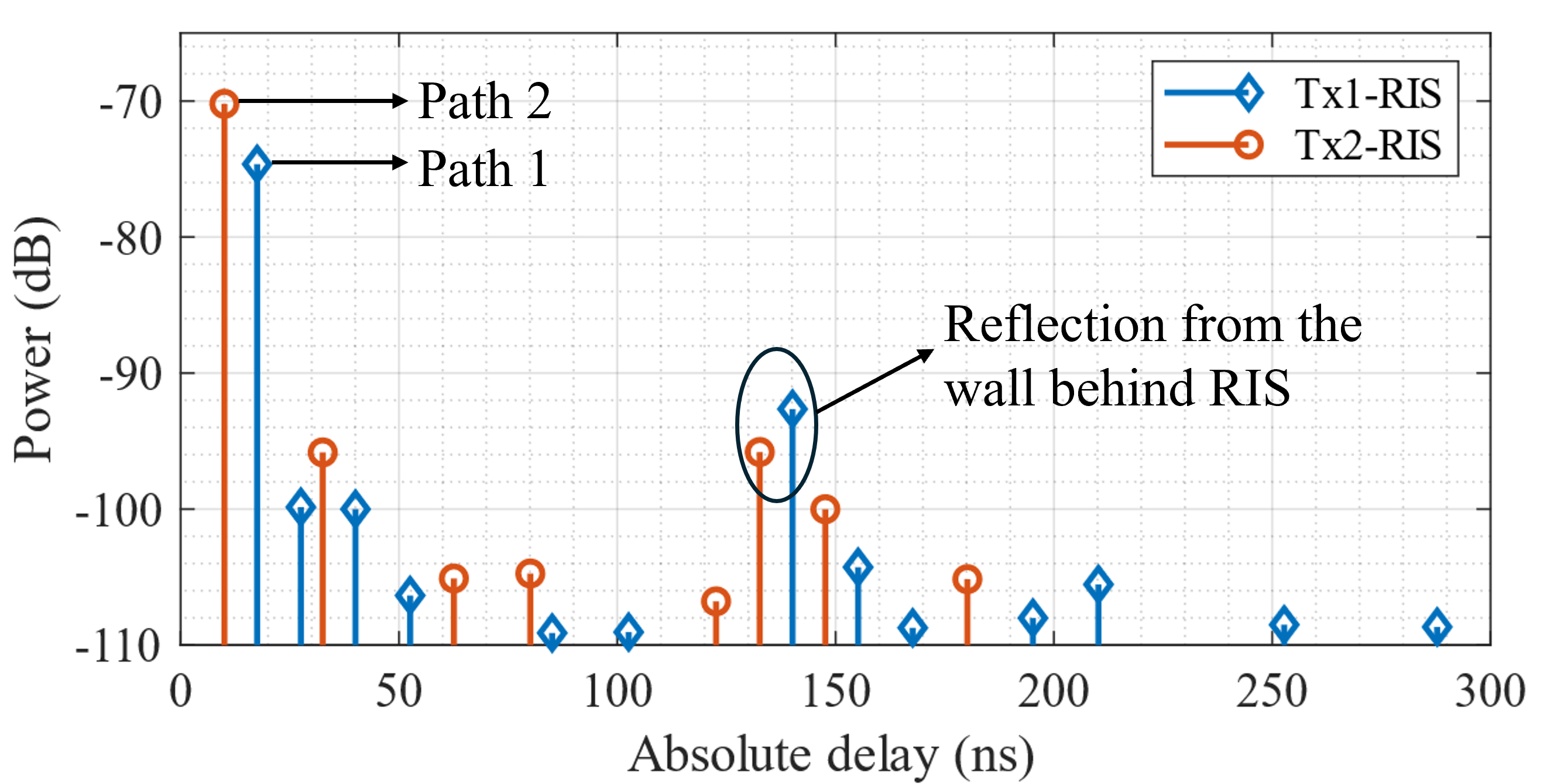}
    \caption{In the effective paths results for the Tx-RIS sub-channel, the blue line and the red line represent the effective paths measured when the transmitting antenna is positioned at Tx1 and Tx2, respectively.}
    \label{txRisPdp}
\end{figure}
\begin{figure*}[!htbp]
    \centering
    \subfloat[]{\includegraphics[width=0.7\columnwidth]{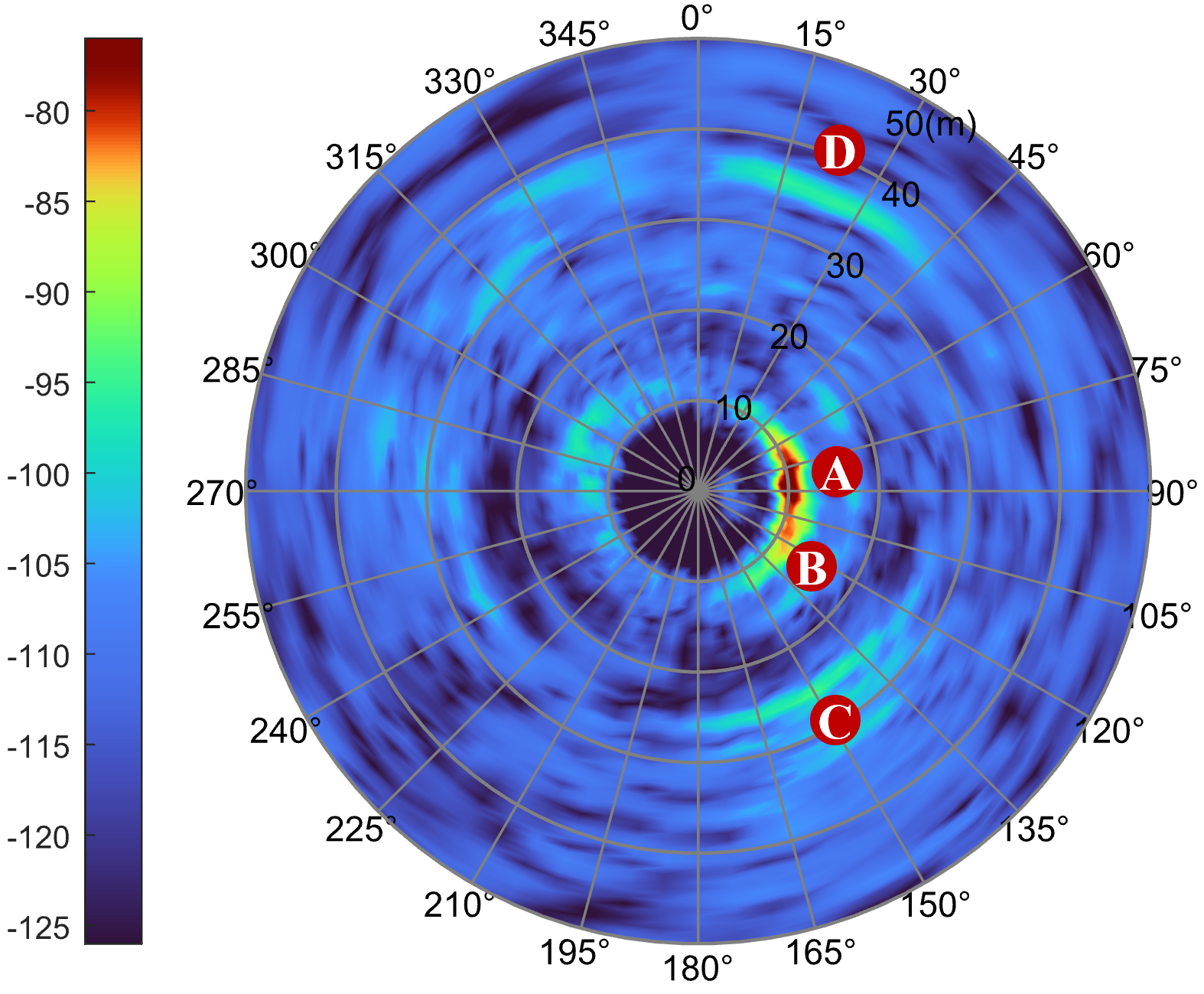}\label{rxPadp}}
    \hfill
    \subfloat[]{\includegraphics[width=0.7\columnwidth]{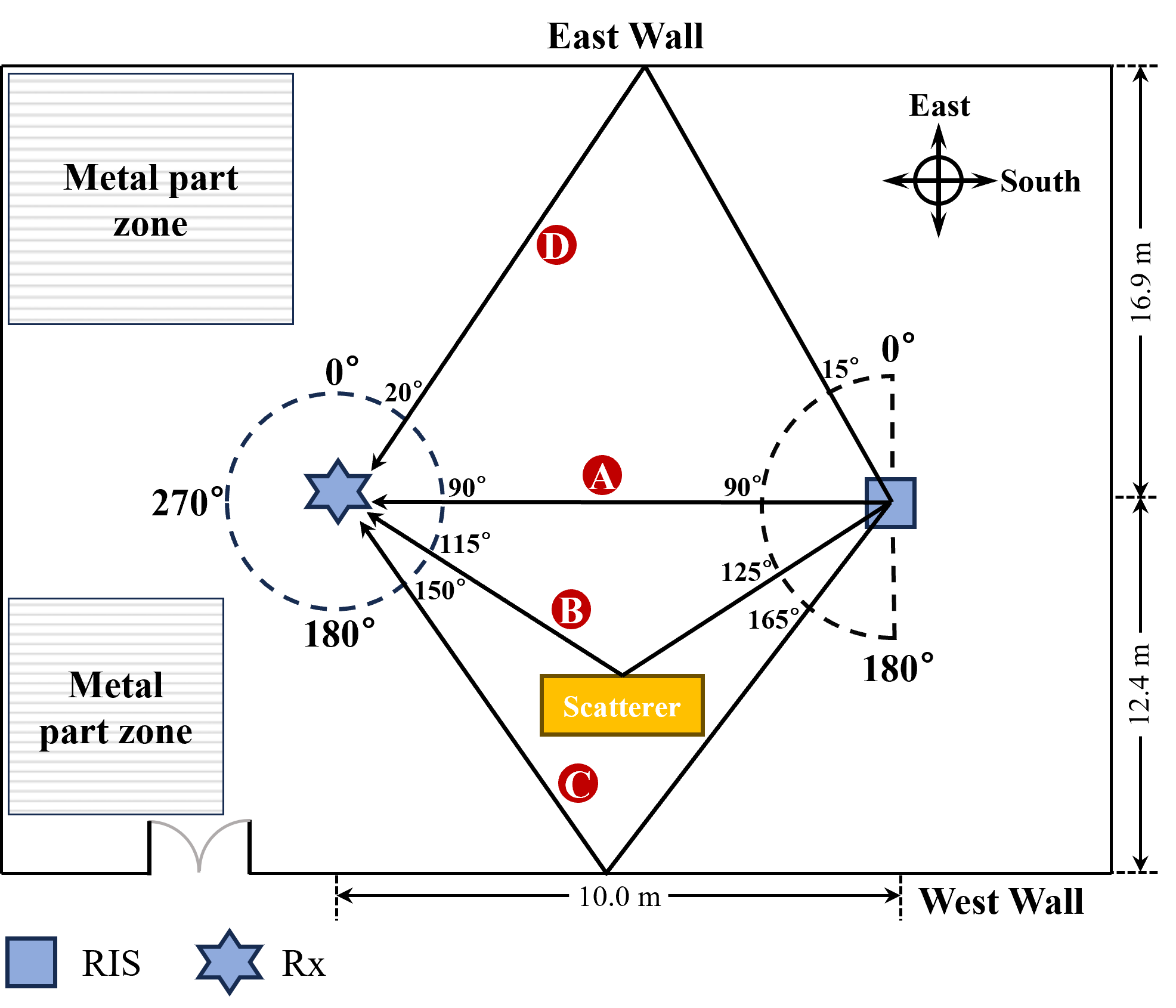}\label{multipathABCD}}
    \hfill
    \subfloat[]{\includegraphics[width=0.42\columnwidth]{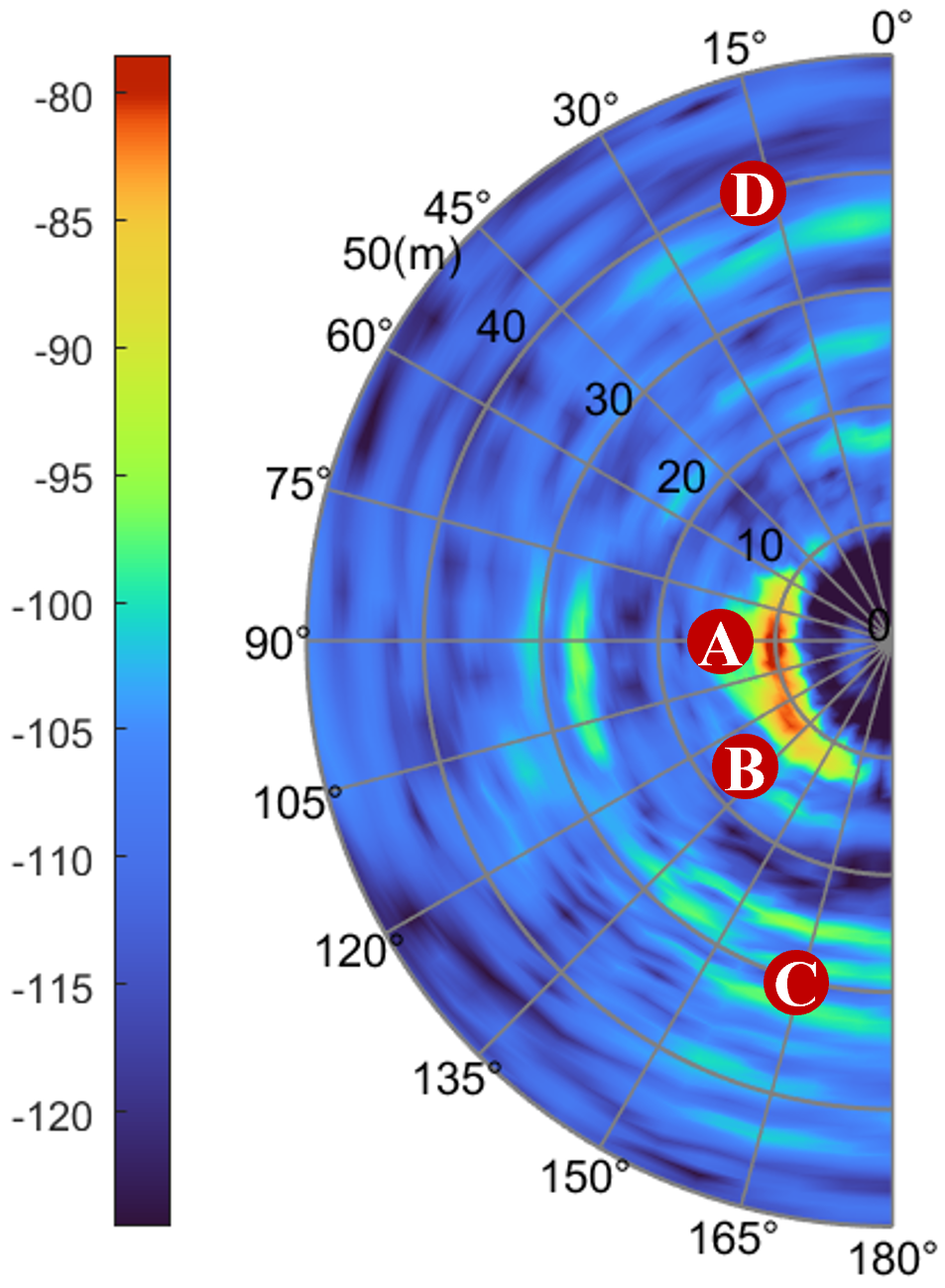}\label{risPadp}}
    \caption{The measurement results for the RIS-Rx sub-channel are as follows: (a) The Power-Angular-Delay Profiles (PADP) at the Rx, where four distinct Path are observed. (b) The possible paths for Path A to D. (c) The PADP at the RIS.}
    \label{RIS-Rx_mul}
\end{figure*}

\section{Measurement Results and Analysis}
In this section, to verify whether the measured data aligns with the RIS convolution cascaded model, the measurement results of the Tx-RIS, RIS-Rx, and Tx-RIS-Rx channel are analyzed sequentially and compared with the theoretical model. Since the analysis is performed at the path level, the path subscript in the model can be simplified from $(n_1, m_1)^{th}$ to $n_1^{th}$.

In the experimental validation, the antenna gain is already removed, and the multipath power can be expressed as
\begin{align}
P_{n}=|h_{\text{RIS}}(\tau_{n})|^2,
\tag{6}
\end{align}
where \( h_{\text{RIS}} \) represents the measured RIS channel CIR, \( P_{n} \), \( \tau_{n} \) represent the power and delay of \( n^{th} \) path in RIS channel, respectively. The sub-channel path power \( P_{n_1} \) and \( P_{n_2} \) can be obtained from the measured CIR in a similar manner.

The theoretical cascade path power can be calculated using (5) as
\begin{align}
P^{\text{conv}}_{n_1,n_2}=P_{n_1}+P_{n_2}+F_{\text{RIS}}(\theta_{n_2}^{\text{out}}, \theta_{n_1}^{\text{in}}),
\tag{7}
\end{align}
where $P^{\text{conv}}_{n_1,n_2}$  refers to the power calculated through the RIS cascaded channel model in convolution form, and all terms are expressed in the form of power or power gain in dB.

The key to validating the accuracy of the cascaded model lies in analyzing the difference between the theoretical path power and the measured path power.
\begin{align}
\Delta P=P^{\text{conv}}_{n_1,n_2}-P_{n},
\tag{8}
\end{align}
If this difference falls within an acceptable range, the model can be considered to fit the measured data well.

\subsection{Tx-RIS Sub-channel}
The PDP of the Tx-RIS sub-channel is shown in Fig.~\ref{txRisPdp}, where the LOS path and the second strongest path are marked. Upon analyzing the scene, it is determined that the second strongest path is a reflection from the wall behind the RIS. In the RIS channel, this path is obstructed by the RIS, so the analysis primarily focuses on the strongest LOS path. The power, delay, and other parameters of these LOS paths, denoted as Path 1 and 2, are presented in Table~\ref{txRisMul}.

\begin{table}[htbp]
\newcommand{\tabincell}[2]{\begin{tabular}{@{}#1@{}}#2\end{tabular}}
\centering
\caption{Path Parameters of the Tx-RIS Sub-channel}
\renewcommand{\arraystretch}{1.5}
\tabcolsep=0.45cm
\begin{tabular}{c|c|c|c}
\hline
\hline
\textbf{Path} & \textbf{\tabincell{c}{Power \\ $\bm{P_{n_1}}$ (dB)}} & \textbf{\tabincell{c}{Delay \\ $\bm{\tau_{n_1}}$ (ns)}} & \textbf{\tabincell{c}{AoA at RIS \\ $\bm{\theta_{n_1}^{\textbf{in}}}$ (deg)}}\\ \hline
1 & -74.64 & 17.5 & 80 \\ \hline
2 & -70.21 & 10.0 & 60 \\ \hline
\hline
\end{tabular}

\label{txRisMul}
\end{table}
\begin{table}[htbp]
\newcommand{\tabincell}[2]{\begin{tabular}{@{}#1@{}}#2\end{tabular}}
\centering
\caption{Path Parameters of the RIS-Rx Sub-channel}
\renewcommand{\arraystretch}{1.5}
\small
\tabcolsep=0.18cm
\begin{tabular}{c|c|c|c|c}
\hline
\hline
\textbf{Path} & \textbf{\tabincell{c}{Power \\ $\bm{P_{n_2}}$ (dB)}} & \textbf{\tabincell{c}{Delay \\ $\bm{\tau_{n_2}}$ (ns)}} & \textbf{\tabincell{c}{AoD at RIS \\ $\bm{\theta_{n_2}^{\textbf{out}}}$ (deg)}} & \textbf{\tabincell{c}{AoA at Rx \\ $\bm{\theta_{n_2}^{\textbf{Rx}}}$ (deg)}}\\ 
\hline
A & -78.46 & 32.5 & 90 & 90\\ \hline
B & -83.36 & 37.5 & 125 & 115\\ \hline
C & -95.59 & 87.5 & 165 & 150\\ \hline
D & -93.28 & 120.0 & 15 & 20\\ \hline
\hline
\end{tabular}
\label{risRxMul}
\end{table}

\begin{figure}[htbp]
    \centering
    \subfloat[]{\includegraphics[width=0.7\columnwidth]{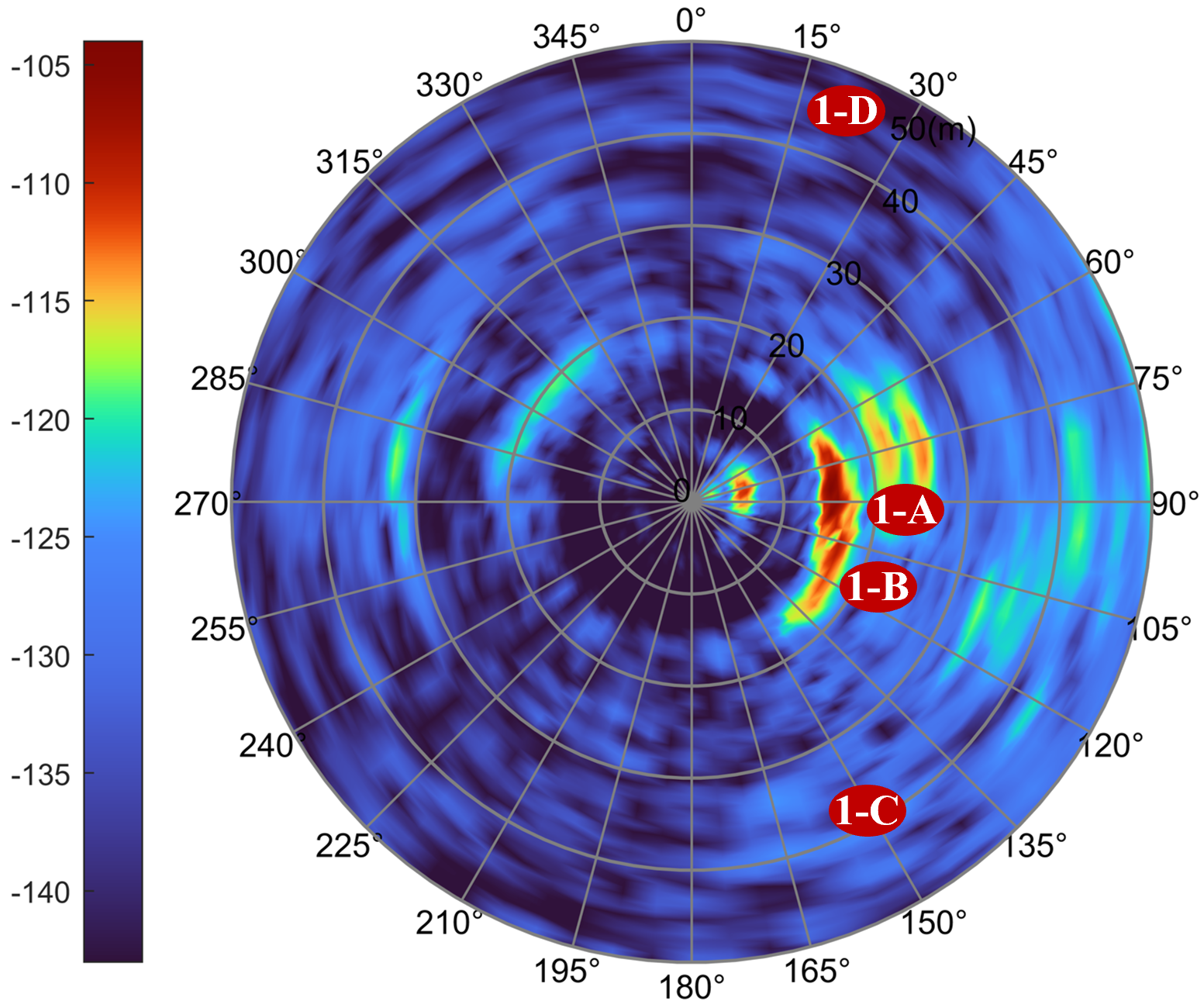}\label{RisDep}}
    \hfill
    \subfloat[]{\includegraphics[width=0.7\columnwidth]{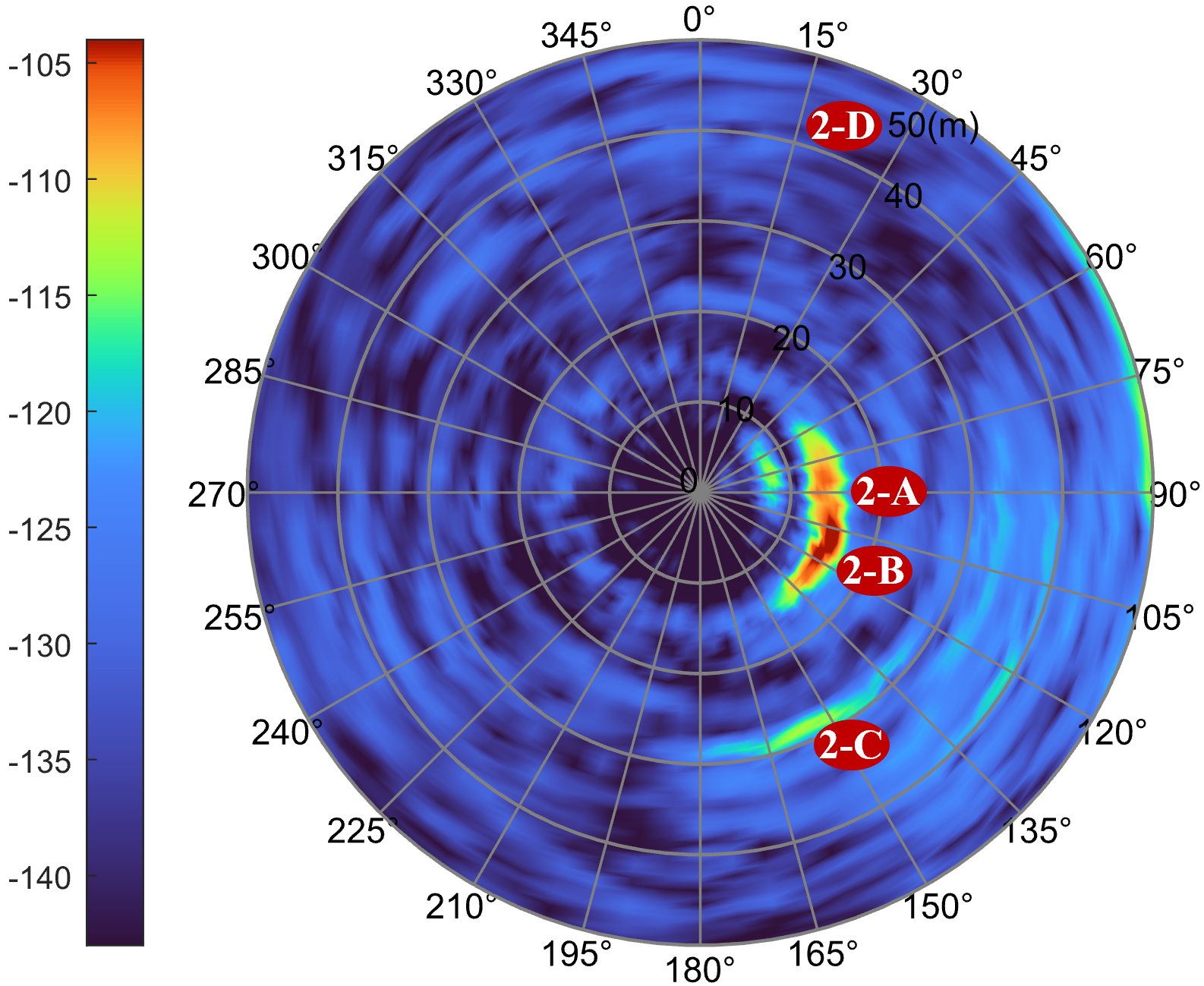}\label{RxArr}}
    \caption{The measured PADP results for the cascaded channel when the RIS is configured to codebook2, (a) with the transmitter located at Tx1, (b) with the transmitter located at Tx2.}
    \label{c2Mul}
\end{figure}
\addtolength{\topmargin}{0.04in}
\begin{table*}[h]
\centering
\newcommand{\tabincell}[2]{\begin{tabular}{@{}#1@{}}#2\end{tabular}}
\caption{Comparison of Results Under Codebook 2}
\renewcommand{\arraystretch}{1.5}
\small
\tabcolsep=0.3cm
\begin{tabular}{c|c|c|c|c|c|c|c}
\hline
\hline
\textbf{Path} & \textbf{ $\bm{ P_{n_1} }$ (dB)} & \textbf{ $\bm{ P_{n_2} }$ (dB)} & \textbf{$\bm{ F_\text{RIS} }$ (dB)} &  \textbf{\tabincell{c}{Cascaded Power \\ $\bm{ P^{\text{conv}}_{n_1,n_2} }$ (dB)}} & \textbf{\tabincell{c}{Actual Received \\ Power $\bm{ P_{n} }$  (dB)}} & \textbf{\tabincell{c}{ Received Power \\ without RIS (dB)}} & \textbf{\tabincell{c}{Difference \\ $\bm{ \Delta P }$  (dB)}} \\ \hline
1-A & -74.64 & -78.46 & 46.71 & -106.39 & -107.54 & -135.38 & 1.15 \\ \hline
2-A & -70.21 & -78.46 & 47.27 & -101.40 & -106.14 & -150.48 & 4.74 \\ \hline
1-B & -74.64 & -83.36 & 52.42 & -105.58 & -112.18 & -151.06 & 6.60 \\ \hline
2-B & -70.21 & -83.36 & 42.69 & -110.88 & -105.81 & -142.97 & -5.07 \\ \hline
1-C & -74.64 & -93.28 & 38.98 & -128.94 & -128.83 & -141.26 & -0.11 \\ \hline
2-C & -70.21 & -93.28 & 44.93 & -118.56 & -113.47 & -147.06 & -5.09 \\ \hline
1-D & -74.64 & -95.59 & 38.69 & -131.54 & -132.19 & -143.43 & 0.65 \\ \hline
2-D & -70.21 & -95.59 & 31.90 & -133.90 & -134.53 & -139.08 & 0.63 \\ \hline
\hline
\end{tabular}
\label{compare}
\end{table*}

\subsection{RIS-Rx Sub-channel}
In the RIS-Rx sub-channel measurement, we obtained the angle of arrival data at the Rx and the angle of departure data at the RIS, as shown in the Fig.~\ref{RIS-Rx_mul}(a), (c). By comparing with the scene diagram, we accurately traced the paths of these four multipath components as shown in Fig.~\ref{RIS-Rx_mul}(b). Path A represents the LOS path between the RIS and Rx, Path B is the reflection path from a deliberately placed scatterer, and Path C and D are the reflection paths from the west and east walls, respectively. The specific parameters corresponding to each path are detailed in Table~\ref{risRxMul}.

\begin{figure}[h]
    \centering
    \includegraphics[width=0.9\columnwidth]{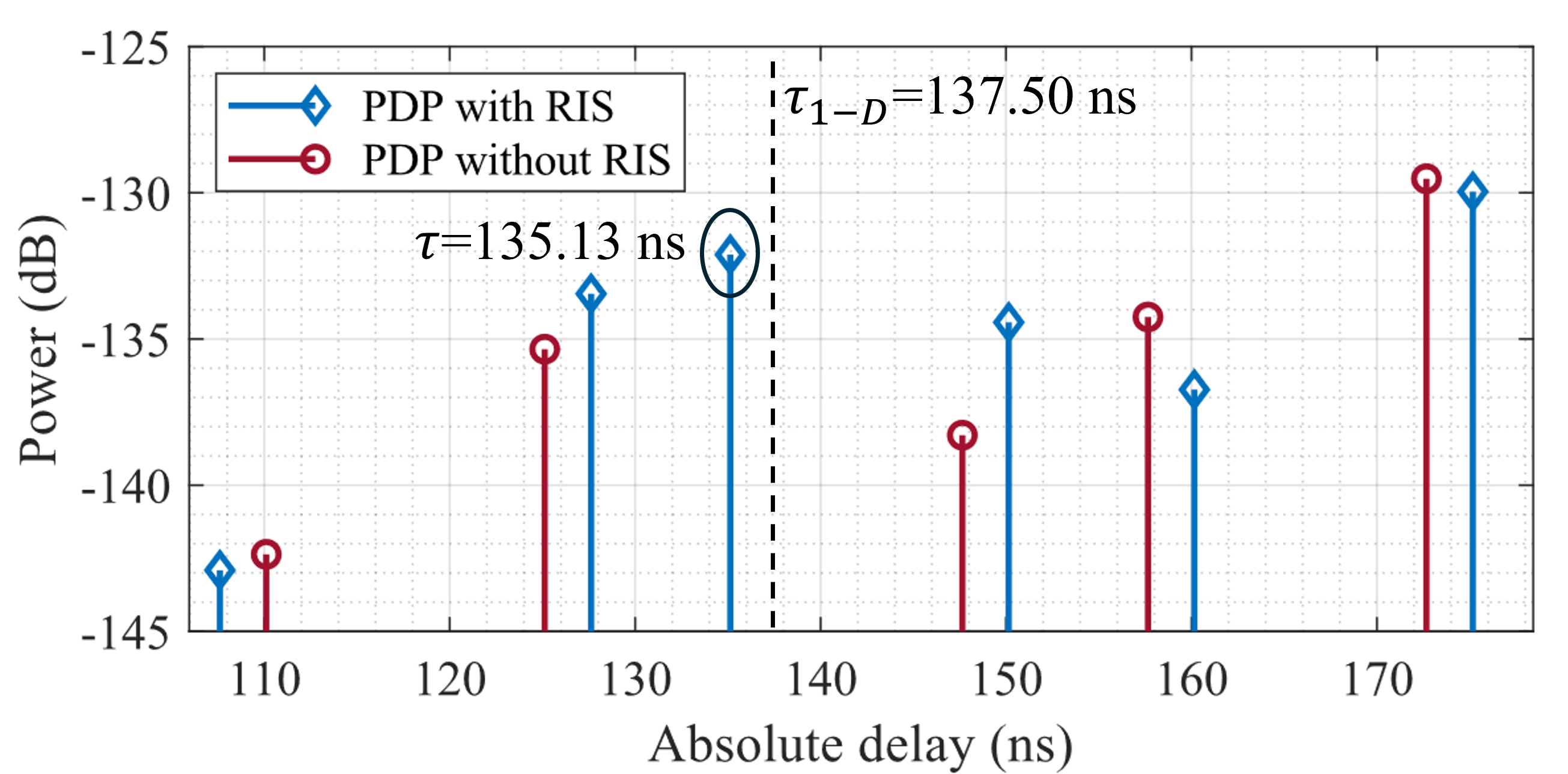}
    \caption{The effective paths of PDP measured when the transmitter is located at position Tx1, and the receiving antenna is oriented at 20 degrees, corresponding to the Rx arrival angle of Path 1-D.}
    \label{c2_1-D_PDP}
\end{figure}

\subsection{Tx-RIS-Rx Cascaded Channel}
Taking the measured PADP under codebook2 as an example, the results are shown in Fig.~\ref{c2Mul}. The parameters of the sub-channel paths are used to deduce the delay and arrival angles for the cascaded paths. ``1-A" indicates that this path is formed by the cascade of Path 1 and A. To obtain more accurate data, we analyze the PDP at the corresponding angles. Taking Path 1-D as an example, as illustrated in Fig.~\ref{c2_1-D_PDP}, it can be observed that, compared to the channel without the RIS, the presence of the RIS introduces new path at the corresponding delays. 

The comparison results are summarized in Table~\ref{compare}.
By comparing the calculated cascaded power with the actual measured power, it is observed that the difference does not exceed 7dB, with the smallest being only 0.11dB.
Considering an angular measurement interval of 5 degrees, which leads to a maximum angular error of 2.5 degrees at \(\theta^\text{out}_{n_2}\), and given the significant fluctuation range of the RIS's radiation pattern, a certain level of difference is considered acceptable.

\section{Conclusion}
Based on the previously proposed GBSM-based RIS channel model, this paper presents a convolutional form RIS cascaded channel model. To verify the validity of this model, measurements are conducted on the RIS channel, the Tx-RIS subchannel, and the RIS-Rx subchannel at a frequency of 6.9 GHz in a factory environment. The model shows a high degree of consistency with the measured results in terms of parameters such as path delay, angle, and power, indicating that this model can be effectively applied to complex real-world scenarios. This provides an important reference for the practical deployment of RIS assisted communication systems.

\section*{Acknowledgment}
This research is supported by National Key R\&D Program of China (2023YFB2904805), Young Scientists Fund of the National Natural Science Foundation of China (62101069, 62201087), Guangdong Major Project of  Basic and Applied Basic Research (2023B0303000001), Key Program of National Natural Science Foundation of China (92167202), National Science Fund for Distinguished Young Scholars (61925102), National Natural Science Foundation of China (62341128) and BUPT-CMCC Joint Innovation Center.

\bibliographystyle{IEEEtran}
\bibliography{references}

\begin{thebibliography}{10}
\providecommand{\url}[1]{#1}
\csname url@samestyle\endcsname
\providecommand{\newblock}{\relax}
\providecommand{\bibinfo}[2]{#2}
\providecommand{\BIBentrySTDinterwordspacing}{\spaceskip=0pt\relax}
\providecommand{\BIBentryALTinterwordstretchfactor}{4}
\providecommand{\BIBentryALTinterwordspacing}{\spaceskip=\fontdimen2\font plus
\BIBentryALTinterwordstretchfactor\fontdimen3\font minus \fontdimen4\font\relax}
\providecommand{\BIBforeignlanguage}[2]{{%
\expandafter\ifx\csname l@#1\endcsname\relax
\typeout{** WARNING: IEEEtran.bst: No hyphenation pattern has been}%
\typeout{** loaded for the language `#1'. Using the pattern for}%
\typeout{** the default language instead.}%
\else
\language=\csname l@#1\endcsname
\fi
#2}}
\providecommand{\BIBdecl}{\relax}
\BIBdecl

\bibitem{ctj}
T.~J. Cui, M.~Q. Qi, X.~Wan, J.~Zhao, and Q.~Cheng, ``Coding metamaterials, digital metamaterials and programmable metamaterials,'' \emph{Light: science \& applications}, vol.~3, no.~10, pp. e218--e218, 2014.

\bibitem{basar}
E.~Basar, M.~Di~Renzo, J.~De~Rosny, M.~Debbah, M.-S. Alouini, and R.~Zhang, ``Wireless communications through reconfigurable intelligent surfaces,'' \emph{IEEE Access}, vol.~7, pp. 116\,753--116\,773, 2019.

\bibitem{zongshu}
\BIBentryALTinterwordspacing
J.~Zhang, J.~Lin, P.~Tang, Y.~Zhang, H.~Xu, T.~Gao, H.~Miao, Z.~Chai, Z.~Zhou, Y.~Li, H.~Gong, Y.~Liu, Z.~Yuan, L.~Tian, S.~Yang, L.~Xia, G.~Liu, and P.~Zhang, ``Channel measurement, modeling, and simulation for 6g: A survey and tutorial,'' 2024. [Online]. Available: \url{https://arxiv.org/abs/2305.16616}
\BIBentrySTDinterwordspacing

\bibitem{mtx1}
Q.~Wu and R.~Zhang, ``Intelligent reflecting surface enhanced wireless network via joint active and passive beamforming,'' \emph{IEEE Transactions on Wireless Communications}, vol.~18, no.~11, pp. 5394--5409, 2019.

\bibitem{mtx2}
Z.~Ding, R.~Schober, and H.~V. Poor, ``On the impact of phase shifting designs on \uppercase{IRS-NOMA},'' \emph{IEEE Wireless Communications Letters}, vol.~9, no.~10, pp. 1596--1600, 2020.

\bibitem{wcx}
J.~Huang, C.-X. Wang, S.~Yang, Y.~Wang, Y.~Xu, Y.~Sun, J.~Huang, and F.-C. Zheng, ``Ray tracing based {6G RIS-Assisted MIMO} channel modeling and verification,'' in \emph{2023 IEEE/CIC International Conference on Communications in China (ICCC)}, 2023, pp. 1--6.

\bibitem{zzf}
J.~Zhang, Z.~Zhou, Y.~Zhang, L.~Tian, Z.~Yuan, and T.~Jiang, ``A deterministic channel modeling method for ris-assisted communication in sub-thz frequencies,'' 2023, accept by 2023 17th European Conference on Antennas and Propagation (EuCAP).

\bibitem{twk}
W.~Tang, M.~Z. Chen, X.~Chen, J.~Y. Dai, Y.~Han, M.~Di~Renzo, Y.~Zeng, S.~Jin, Q.~Cheng, and T.~J. Cui, ``Wireless communications with reconfigurable intelligent surface: Path loss modeling and experimental measurement,'' \emph{IEEE Transactions on Wireless Communications}, vol.~20, no.~1, pp. 421--439, 2021.

\bibitem{ly}
Y.~Li, J.~Zhang, P.~Tang, L.~Tian, X.~Zhao, H.~Xu, and H.~Gong, ``Path loss modeling for the {RIS-Assisted} channel in a corridor scenario in mmwave bands,'' in \emph{2022 IEEE Globecom Workshops (GC Wkshps)}, 2022, pp. 1478--1483.

\bibitem{shibie}
M.~El-Absi, A.~A. Abbas, D.~Tubail, F.~Ilgac, A.~Abuelhaija, Y.~Zantah, S.~Ikki, A.~Sezgin, and T.~Kaiser, ``Path loss modeling of rfid backscatter channels with reconfigurable intelligent surface: Experimental validation,'' \emph{IEEE Access}, vol.~11, pp. 108\,532--108\,543, 2023.

\bibitem{ghw}
H.~Gong, J.~Zhang, Y.~Zhang, Z.~Zhou, and G.~Liu, ``How to extend {3-D GBSM to RIS} cascade channel with non-ideal phase modulation?'' \emph{IEEE Wireless Communications Letters}, vol.~13, no.~2, pp. 555--559, 2024.

\bibitem{cluster}
J.~Zhang, C.~Pan, F.~Pei, G.~Liu, and X.~Cheng, ``Three-dimensional fading channel models: A survey of elevation angle research,'' \emph{IEEE Communications Magazine}, vol.~52, no.~6, pp. 218--226, 2014.

\bibitem{38901}
\BIBentryALTinterwordspacing
3GPP, ``{Study on Channel Model for Frequencies from 0.5 to 100 GHz.}'' {3rd Generation Partnership Project (3GPP)}, Technical Specification (TR) 38.901, 03 2022, version 17.0.0. [Online]. Available: \url{www.3gpp.org}
\BIBentrySTDinterwordspacing

\end{thebibliography}

\end{document}